\documentclass[prd,aps,showpacs,twocolumn,tightenlines,superscriptaddress,amsmath,amssymb,amsfonts]{revtex4-2}
\usepackage{amssymb,amsmath,amsthm,amsbsy,epsfig,color,graphicx,times,physics}
\usepackage{appendix}

\usepackage[ansinew]{inputenc}
\usepackage[english]{babel}
\usepackage{xcolor}

\usepackage{braket}

\usepackage{tabularx}
\usepackage{array}

\usepackage{slashed}
\usepackage{subcaption}

\newcommand{\blue}[1]{\textcolor{blue}{#1}}

\begin{document}

\title{A Breakdown Case Study of the Lindblad Approach via Entanglement and Purity}

\author{R. Serao}
\email{rserao@unisa.it}
\affiliation{Dipartimento di Fisica ``E.R. Caianiello'' Universit\`{a} di Salerno, and INFN --- Gruppo Collegato di Salerno, Via Giovanni Paolo II, 132, 84084 Fisciano (SA), Italy}

\author{A. Quaranta}
\email{anquaranta@sa.infn.it}
\affiliation{School of Science and Technology, University of Camerino, Via Madonna delle Carceri, Camerino, 62032, Italy}

\author{A. Capolupo}
\email{capolupo@sa.infn.it}
\affiliation{Dipartimento di Fisica ``E.R. Caianiello'' Universit\`{a} di Salerno, and INFN --- Gruppo Collegato di Salerno, Via Giovanni Paolo II, 132, 84084 Fisciano (SA), Italy}

\author{F. Franchini}
\email{fabio.franchini@irb.hr}
\affiliation{Institut Ru\dj er Bo\v{s}kovi\'c, Bijeni\v{c}ka cesta 54, 10000 Zagreb, Croatia}

\author{S.M. Giampaolo}
\email{sgiampa@irb.hr}
\affiliation{Institut Ru\dj er Bo\v{s}kovi\'c, Bijeni\v{c}ka cesta 54, 10000 Zagreb, Croatia}

\begin{abstract}
\blue{We investigate the reduced dynamics of two interacting two-level subsystems coupled to a finite many-body environment through a pure-dephasing interaction that commutes with both the system and environment Hamiltonians. For strongly correlated couplings, the exact dynamics exhibits two well-separated timescales: rapid collective dephasing, followed by slower decoherence driven by relative environmental fluctuations. Both regimes display approximately Gaussian coherence decay, which cannot be reproduced by the exponential factors of the time-homogeneous pure-dephasing Lindblad model considered here.}
\end{abstract}

\maketitle

\section{Introduction}

The interaction between a quantum system and its surrounding environment is widely recognized as a fundamental mechanism responsible for decoherence and the emergence of classical behavior in quantum systems~\cite{Zurek2003, Breuer2007}.
With the rapid development of quantum information science, understanding and accurately predicting how environmental degrees of freedom affect different quantum resources has become increasingly crucial.
In particular, the dynamical behavior of quantum resources~\cite{Chitambar2019} such as entanglement~\cite{Bennet1996, Horodecki2009}, coherence~\cite{Baumgratz2014, Streltsov2017} non-stabilizerness~\cite{Bravyi2006, Veitch2014} (magic), as well as quantities such as purity~\cite{Nielsen2010, Bengtsson2017}, plays a central role in assessing the performance and robustness of quantum technologies operating in realistic open-system settings.
Clarifying how these resources degrade, persist, or transform under environmental interactions therefore represents a fundamental challenge across several areas of quantum physics, ranging from quantum information processing to condensed matter physics and quantum foundations.

A standard theoretical framework to describe open quantum systems is provided by the Gorini--Kossakowski--Sudarshan--Lindblad (GKSL) master equation~\cite{Gorini1976, Lindblad1976}.
Within this approach, the reduced dynamics of the system is assumed to be Markovian and time-homogeneous, leading to a quantum dynamical semigroup that guarantees complete positivity and trace preservation~\cite{Breuer2007}.
This framework has been successfully applied to a wide variety of physical scenarios and is often employed as an effective description of environmental effects~\cite{Rivas2012, Vega2017,Stefanini2025}.

To provide an example, the GKSL formalism has also found important applications in the context of fundamental particle physics, where open-system techniques have been employed to model possible deviations from standard unitary dynamics.
In a series of seminal works, Benatti and Floreanini investigated neutrino oscillations and neutral meson systems within a completely positive open quantum system framework~\cite{BE1, BE2, BE3, BE4, BE5}.
In these studies, Lindblad-type master equations were used to parametrize decoherence and dissipation effects arising from hypothetical environmental or quantum-gravity interactions, while preserving fundamental consistency requirements such as complete positivity.
These works provided a systematic and phenomenologically consistent approach to incorporating environmental effects in oscillation phenomena, and played a key role in establishing open-system methods as viable tools in high-energy and particle physics.
Building on this approach, a broad body of subsequent literature~\cite{Fogli2007, Capolupo2019} has explored the role of decoherence in particle physics, investigating its impact on neutrino oscillations, meson systems, and possible signatures of physics beyond the standard unitary framework.

Despite its broad applicability, the Lindblad approach relies on assumptions, such as weak system-environment coupling, short environmental correlation times, and the absence of persistent system-environment correlations, that are not always satisfied in microscopic models~\cite{Breuer2007, Rivas2014}.
In particular, when the environment has an internal structure or when the system--environment coupling induces correlations that persist over time, non-Markovian effects may arise~\cite{Breuer2009,  Rivas2010}.
In such cases, it is not a priori clear whether a time-homogeneous GKSL generator can faithfully reproduce the reduced dynamics emerging from an underlying unitary evolution.

\blue{Pure-dephasing models provide a paradigmatic setting in which these issues have been extensively investigated. 
Their study has a long history, ranging from early analyses of the pure dephasing of two-level systems to nonperturbative treatments of decoherence in condensed phases~\cite{SkinnerHsu1986, Reichman1996}. 
In particular, the non-exponential decay of quantum coherences, including a Gaussian behavior at short times, is a well-known feature of this class of models~\cite{Reichman1996, Paladino2014, Paganelli2002}. 
Closely related behavior also appears in central-spin models, where exact or nonperturbative treatments have shown that the reduced dynamics can strongly depart from the predictions of the Born--Markov approximation and of Markovian master equations~\cite{Breuer2004, Barnes2012}. 
These results make clear that neither Gaussian short-time dephasing nor the failure of a Markovian approximation is, by itself, peculiar to the microscopic model considered in the present work.
Pure dephasing has also been investigated for bipartite systems, although this setting has received comparatively less attention. 
In particular, Kwiatkowski and Cywi\'nski studied the pure dephasing of two entangled spin qubits interacting with a common nuclear-spin environment, explicitly analyzing the role of correlated environmental fluctuations~\cite{Kwiatkowski2018}. 
More generally, genuinely two-qubit phase-damping dynamics and two-qubit probes undergoing common-bath pure dephasing have also been considered in Refs.~\cite{HelmStrunz2009, Gebbia2020}. 
Accordingly, the purpose of the present work is not to identify Gaussian short-time dephasing, or the failure of a Markovian approximation in a generic pure-dephasing model, as new phenomena.
Rather, our aim is to use an exactly solvable finite many-body environment to isolate two parametrically separated decoherence channels of a bipartite subsystem, associated respectively with collective and relative environmental fluctuations, and to determine explicitly whether their reduced dynamics can be represented by a single time-homogeneous GKSL semigroup.}

Recent interest in this problem has been driven by physical scenarios in which entanglement is generated through coherent interactions with an environment made of a large number of quantum objects, rather than through dissipative processes.
Representative examples include gravitationally induced entanglement between massive systems~\cite{Bose2017, Marletto2017, Giampaolo2019}, particle mixing phenomena~\cite{Capolupo2021}, and many-body systems with long-range interactions~\cite{Paganelli2002, Simonov2019}.
In these settings, the loss of coherence does not arise from energy exchange with the environment, but from dephasing induced by coherent superpositions of distinct unitary evolutions.
The resulting dynamics is intrinsically non-dissipative and reflects the build-up of system--environment correlations at the microscopic level.
This qualitative difference raises a fundamental question: to what extent such dephasing-driven decoherence mechanisms can be faithfully described by a time-homogeneous Markovian Lindblad description.

In this work, we address this issue by analyzing a fully unitary microscopic model in which two interacting two-level subsystems are embedded in a larger many-body environment and coupled through pairwise interactions.
\blue{Our focus is specifically on the bipartite nature of the subsystem and on how the presence of a common many-body environment gives rise to distinct decoherence channels and dynamical timescales.}

The simplicity of the model allows for an exact treatment of the reduced dynamics, making it possible to identify distinct dynamical regimes and to characterize the corresponding decay of coherence, purity, and entanglement.
\blue{In particular, when the fluctuations experienced by the two subsystems are strongly correlated, the reduced dynamics exhibits a clear separation of timescales, giving rise to a fast regime dominated by collective decoherence and a slower regime governed by relative environmental fluctuations.}

We then attempt to reproduce both these regimes within an effective Lindblad framework.
By systematically comparing the exact microscopic evolution with the predictions of a time-homogeneous GKSL master equation, we demonstrate that, although the Lindblad approach can be phenomenologically tuned to reproduce which matrix elements decay in each regime, it fails to capture the correct functional form of the decay.
In particular, the Gaussian suppression of coherence emerging from the microscopic dynamics is fundamentally incompatible with the exponential behavior enforced by the semigroup structure of the Lindblad equation.

\blue{The model has non-decaying environmental correlations and therefore does not satisfy the short-correlation-time assumption of the usual Born--Markov derivation. 
The disagreement with a time-homogeneous semigroup is consequently expected. 
Our contribution is the explicit comparison of the collective and relative dephasing channels in this bipartite finite-bath setting, rather than a general failure of the GKSL framework.}

\section{The isolated two-qubit model}

We begin by considering a bipartite quantum system consisting of two interacting subsystems, denoted by $A$ and $B$, each described by a two-level Hilbert space.
We assume that the Hamiltonian governing the dynamics of the composite system can be written in the form
\begin{equation}\label{HAB}
H_{AB}=\sum_{ij} g_{ij}\ket{a_i b_j}\!\bra{a_i b_j}.
\end{equation}
Within this framework, the eigenstates of $H_{AB}$ are tensor products of states associated with the individual subsystems, namely $\ket{a_i}\otimes\ket{b_j}$.
Here, $\{\ket{a_1},\ket{a_2}\}$ label the basis states of subsystem $A$, while $\{\ket{b_1},\ket{b_2}\}$ label those of subsystem $B$.

Despite its simplicity, this minimal setting already captures the essential features of interaction-induced quantum correlations and therefore provides a natural starting point for our analysis.
At the same time, the model is remarkably general and applies to a broad class of physically relevant scenarios.
A notable example is provided by the BMV experimental proposal~\cite{Bose2017, Marletto2017, Giampaolo2019}, which aims to probe the quantum nature of gravity~\cite{Kiefer2005, Rovelli2008, Oriti2009, Blau2009, Hossenfelder2011, Ashoorioon2014}.
In that context, the basis states $\{\ket{a_i},\ket{b_j}\}$ correspond to different spatial configurations of the interacting systems.
More generally, the same formalism also applies to internal degrees of freedom, such as spin~\cite{Capolupo2020}, as well as to mass eigenstates in particle-mixing phenomena~\cite{Marletto2018}.

Introducing a Pauli operator algebra in each single system and defining 
$\sigma_A^z=\ket{a_1}\bra{a_1}-\ket{a_2}\bra{a_2}$, on $A$ and, analogously, 
$\sigma_B^z=\ket{b_1}\bra{b_1}-\ket{b_2}\bra{b_2}$ on $B$, the Hamiltonian in Eq.~\eqref{HAB} can be rewritten as
\begin{equation}\label{HAB_1}
H_{AB} \!=\!c_0 \sigma^0_{A} \!\otimes\! \sigma^0_{B}\!+\! c_A \sigma^z_A \!\otimes \! \sigma^0_{B}\!+\! c_B \sigma^0_{A} \! \otimes \! \sigma^z_B\!-\!\omega \sigma^z_A \! \otimes \! \sigma^z_B
\end{equation}
Here, $\sigma^0_\alpha$ with $\alpha=\ a,\ b$ denotes the identity operator, and the coefficients $c_0$, $c_A$, $c_B$, and $\omega$ are linear combinations of the parameters $g_{ij}$.

In principle, all terms in the Hamiltonian in Eq.~\eqref{HAB_1} contribute to the unitary time evolution of the system.
However, since the present work focuses on quantities such as entanglement and purity, the analysis can be significantly simplified.
Indeed, all four terms appearing in Eq.~\eqref{HAB_1} commute with one another, which allows the time-evolution operator to be factorized into a product of unitary operators generated by each term separately.
The first contribution is proportional to the identity and yields only a global phase.
The second and third terms generate local unitary dynamics acting exclusively on subsystems $A$ and $B$, respectively.
It is well known that local unitary operations leave both the entanglement between the subsystems and the purity of the state unchanged~\cite{Plenio2007, Nielsen2010}.
Accordingly, these contributions can be disregarded for the purposes of the present analysis.
The relevant dynamics is therefore entirely governed by the remaining nonlocal interaction term:
\begin{equation}
\label{HAB_2}
H_{AB}=-\omega\,\sigma_A^z\otimes\sigma_B^z,
\end{equation}
with $\omega=\tfrac{1}{4}\left(g_{12}+g_{21}-g_{11}-g_{22}\right)$.

The Hamiltonian in Eq.~\eqref{HAB_2} already leads to a nontrivial evolution of the entanglement.
Starting from an initially separable state $\ket{\psi(0)}=\ket{\psi_A}\ket{\psi_B}$, the unitary evolution generates entanglement whenever the local states $\ket{\psi_A}$ and $\ket{\psi_B}$ are prepared as coherent superpositions of the basis states $\ket{a_i}$ and $\ket{b_j}$.
In this case, the interaction term induces correlations between the two subsystems, resulting in a finite amount of entanglement at later times.

\blue{In the case of the interaction Hamiltonian in Eq.~\eqref{HAB_2}, the amount of entanglement generated from an initially separable state depends on the local preparation of the two subsystems. 
Assuming arbitrary normalized pure initial states $\ket{\psi_A}=u_1\ket{a_1}+u_2\ket{a_2}$ and $\ket{\psi_B}=v_1\ket{b_1}+v_2\ket{b_2}$, the concurrence is
\begin{equation}
C_{\mathrm{prod}}(t) =4|u_1u_2v_1v_2|\,|\sin(2\omega t)| \leq|\sin(2\omega t)|.
\label{eq:product_concurrence}
\end{equation}
Equal-magnitude local amplitudes therefore maximize the concurrence at every fixed time within this class. 
The local phases do not affect this result and, hence, for sake of simplicity, are set to zero. 
Explicitly, we take}
\begin{equation}
\label{initialstate}
\begin{split}
\ket{\psi_A} &= \frac{1}{\sqrt{2}}(\ket{a_1}+\ket{a_2}),\\
\ket{\psi_B} &= \frac{1}{\sqrt{2}}(\ket{b_1}+\ket{b_2}).
\end{split}
\end{equation}
For this reason, we henceforth restrict our attention to this specific choice of initial state.
Under this assumption, the entanglement generated by the unitary dynamics, as quantified by the concurrence~\cite{Hill1997,Wootters1998}, is given by
$|\sin\!\left(2 \omega t\right)|$.
In contrast, since the system is closed and undergoes purely unitary evolution, the purity of the state remains constant in time and is identically equal to unity.

This isolated dynamics will serve as a reference benchmark when the interaction with a many-body environment is introduced in the following sections.

\section{The exact solution of the microscopic model}

How does this picture change when $A$ and $B$ -- which we henceforth collectively denote as the subsystem $AB$ -- are no longer treated as isolated, but instead regarded as subsystems of a larger many-body system in which all constituents interact pairwise through a Hamiltonian of the same form as in Eq.~\eqref{HAB}?
A concrete example of such a scenario is provided in Ref.~\cite{Simonov2019}, where the mixing of gravitationally interacting particles is analyzed.
In that context, the internal states of the particles differ in their spatial position and/or masses.
As a consequence, the interaction strength with the surrounding particles depends on the internal configuration of both the system and the environment.

For a many-body system in which all constituents interact pairwise through a Hamiltonian of the form in Eq.~\eqref{HAB}, the total Hamiltonian can be written as
\begin{equation}\label{FullHamiltonian}
	H = H_{AB} \otimes \mathbb{I}_{E} \!+\! H_{AE} \otimes \sigma^0_{B} \!+\! H_{BE} \otimes \sigma^0_{A} \!+\! H_E \otimes \sigma^0_{A}  \otimes \sigma^0_{B}\, ,
\end{equation}
where $\mathbb{I}_{E}$ is the identity operator defined in the Hilbert space of the environment.
This decomposition makes explicit the different physical roles played by the subsystem $AB$, the environment, and their mutual interactions.
Indeed, the presence of the environment adds three Hamiltonian contributions beyond the term describing the internal dynamics of the subsystem $AB$.
The term $H_E$ defines the Hamiltonian of the environment, accounting for all pairwise interactions among the constituents excluding $A$ and $B$.
The remaining two terms describe the coupling between the two parts of the subsystem $AB$ and the environment.
\blue{Since we are interested in the entanglement between $A$ and $B$, we omit local terms acting on individual constituents. 
Within the assumed diagonal structure, these terms commute with the interactions and contribute only local unitary transformations, which do not affect the entanglement dynamics. 
The system--environment interaction terms can therefore be written as
\begin{align}	
	\label{eq:Ham_int_3}
	H_{AE} &=-\sum_{\ell=1}^{N}\omega_{\ell A}\,
	\sigma_A^z\otimes\sigma_\ell^z=\sigma_A^z \otimes \sum_{k} \chi_k^{a}\ket{e_k}\!\bra{e_k} \\
	H_{BE} &=-\sum_{\ell=1}^{N}\omega_{\ell B}\,
	\sigma_B^z\otimes\sigma_\ell^z=\sigma_B^z \otimes \sum_{k} \chi_k^{b}\ket{e_k}\!\bra{e_k}\,.
\end{align}
Here $\omega_{\ell \alpha}$ is the coupling strength between one of the two elements $\alpha=A,B$ and the $\ell$-th environmental constituent.
On the other hand, the coefficients $\chi_k^\alpha$ quantify the dependence of the system-environment interaction on the internal state of subsystem $\alpha$ and on the environmental configuration. 
In terms of the pairwise coupling strengths introduced above, these coefficients read
	\begin{equation}
		\chi_k^\alpha = \sum_{j=1}^{N} (-1)^{(m_j^k+1)/2} \omega_{j\alpha}\, , 
\end{equation}
where $m_j^k = \langle e_k|\sigma_j^z|e_k\rangle = \pm1$.}

Hence, $\chi_k^\alpha $ is the sum of the frequencies weighted by whether the $j$-th component of the environmental state has the same or the opposite orientation of $\alpha$.
If the dependence on the state of the subsystem vanishes, i.e., if $\omega_{\ell \alpha}=0$ for all $\ell$, then $\chi_k^\alpha=0$.
When this condition holds, the dynamics of $AB$ becomes effectively decoupled from the environment.
On the other hand, if $\chi_k^\alpha= \bar{\chi}^\alpha$ for all $k$ and $\alpha$, the resolution of identity $\sum_k \ket{e_k}\!\bra{e_k}=\mathbb{I}_E$ implies that the interaction terms reduce to local unitary fields acting independently on subsystems $A$ and $B$.
Consequently, they do not modify the evolution of either the purity of the reduced state or the entanglement between the two subsystems.
These limiting cases, however, correspond to finely tuned situations and do not capture the generic behavior arising from state-dependent interactions with a many-body environment, which we analyze in detail in the following.

We now turn to the analysis of the entanglement and purity dynamics and compare them with the behavior observed in the isolated case.
\blue{The system--environment interactions in Eq.~\eqref{eq:Ham_int_3} commute with both $H_{AB}$ and $H_E$, so that the environment induces pure dephasing.}
To fix the ideas, as in the previous scenario, we assume the initial state of the total system, i.e., the subsystems $AB$ and the environment, to be fully factorized,
$\ket{\psi(0)}=\ket{\psi_A}\ket{\psi_B}\ket{\psi_E}$, with $\ket{\psi_A}$ and $\ket{\psi_B}$ chosen as in Eq.~\eqref{initialstate}.
By contrast, the initial pure state of the environment is taken to be generic, i.e.,
\begin{equation}
	\ket{\psi_E}=\sum_k f_k \ket{e_k},
	\label{psieexp}
\end{equation}
where the complex coefficients $f_k$ are subject only to the normalization condition $\sum_k |f_k|^2 = 1$.

Under these assumptions, the reduced density matrix describing the joint state of subsystems $A$ and $B$, obtained after tracing out the environmental degrees of freedom, reads
\begin{equation}
	\label{rho_time_1}
\!\!\!\!\!\rho_{AB}(t)\!=\!\!
	\frac{1}{4}\!\!
	\begin{pmatrix}
		1 & e^{\imath 2 \omega t}\Gamma_B & e^{\imath 2 \omega t}\Gamma_A & \Lambda_+ \\
		e^{-\imath 2 \omega t}\Gamma_B^* & 1 & \Lambda_- & e^{-\imath 2 \omega t}\Gamma_A \\
		e^{-\imath 2 \omega t}\Gamma_A^* & \Lambda_-^* & 1 & e^{-\imath 2 \omega t}\Gamma_B \\
		\Lambda_+^* & e^{\imath 2 \omega t} \Gamma_A^* & e^{\imath 2 \omega t} \Gamma_B^* & 1
	\end{pmatrix}\!\!,
\end{equation}
where the functions appearing in the reduced density matrix are given by
\begin{eqnarray}
	\label{rho_time_1A}
	\Gamma_\alpha &=& \sum_k |f_k|^2 e^{-2 \imath \chi_k^\alpha t}, \nonumber \\
	\Lambda_\pm &=& \sum_k |f_k|^2 e^{-2 \imath (\chi_k^a \pm \chi_k^b)t}.
\end{eqnarray}
\blue{Since the environmental coupling operators commute with $H_E$, their connected correlation functions are time independent. 
The bath correlation time is therefore formally infinite for channels with nonzero fluctuations, and the usual short-correlation-time Born--Markov approximation is not justified.}

From Eq.~\eqref{rho_time_1A}, one may be led to conclude that the quantities $\Gamma_\alpha$ and $\Lambda_\pm$ depend on the intrinsic dynamics of the environment generated by $H_E$.
This conclusion is, however, incorrect.
Since the set of states $\{\ket{e_k}\}$ represents a complete set of eigenstates of 
$H_E$, the coefficients $f_k$ evolve as $f_k(t)=f_k\,e^{-\imath \varepsilon_k t}$, where $\varepsilon_k$ is the eigenvalue of $H_E$ associated with the eigenstate $\ket{e_k}$.
It follows immediately that $|f_k(t)|=|f_k|$ at all times.
As a consequence, the unitary evolution generated by $H_E$ does not influence the dynamics of any physical quantity defined on the subsystem $AB$, including its entanglement properties.
Moreover, since the reduced density matrix depends solely on the moduli of the coefficients $f_k$, it is insensitive to the presence or absence of quantum coherence in the environmental state.
Therefore, once the set of coefficients $|f_k|$ is specified, the same reduced dynamics is obtained whether the environment is prepared in a coherent pure state, as assumed so far, or in \blue{a mixed state with the same diagonal populations}.
However, although the set of coefficients $|f_k|$ is time independent, the functions $\Gamma_\alpha$ and $\Lambda_\pm$ exhibit a nontrivial time dependence.
As a result, they contribute significantly to the evolution of physical quantities such as entanglement and purity, as well as to the overall dynamics of the reduced density matrix $\rho_{AB}(t)$. Before discussing the evolution of these quantities in generality, through analytical arguments, let us focus on a numerical instance to introduce some general concepts.

\begin{figure}[t!]
	\centering
	\hspace{-1.30cm}\includegraphics[width=1.15\columnwidth]{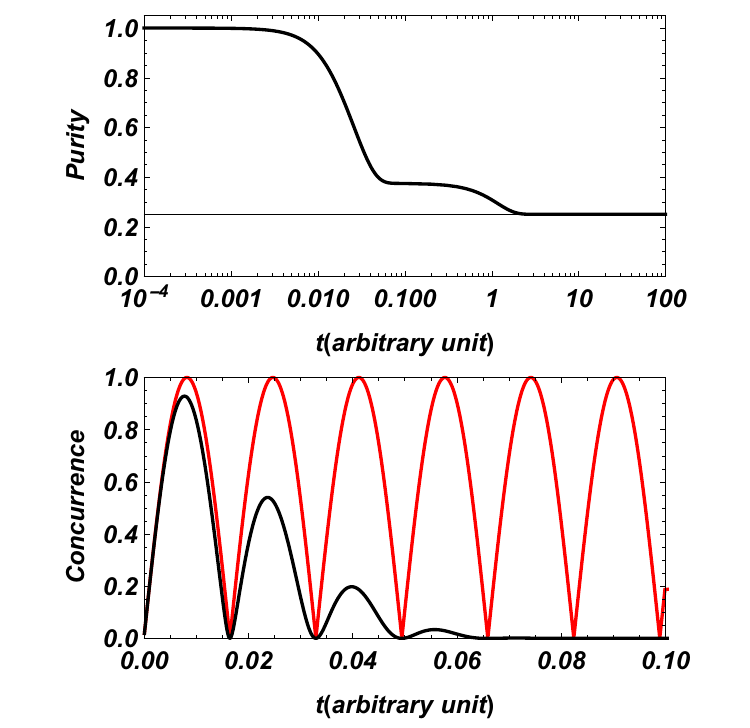}
	\caption{Top panel: Time evolution of the purity of the reduced state $\rho_{AB}(t)$ defined in Eq.~\eqref{rho_time_1}.
		The lower horizontal line at $1/4$ denotes the minimum purity attainable for a system composed of two two-level subsystems.
		The upper horizontal line at $3/8$ indicates the value of the purity of the matrix in Eq.~\eqref{rho_time_lim} that is reached at the end of the first dynamical regime.
		Bottom panel: Time evolution of the concurrence associated with $\rho_{AB}(t)$ (solid black line), compared with the concurrence obtained by neglecting the presence of the environment (red line).
		Both panels correspond to the same physical realization: a fixed number of two-level systems ($N=502$) randomly distributed within a sphere of radius $R=1$ (arbitrary units).
		Each pair of two-level systems interacts through a Hamiltonian of the form given in Eq.~\eqref{HAB_2}, with the coupling strength $\omega_{ij}$ taken to be the inverse of the relative distance.
		The pair with the smallest separation is identified as the subsystem $AB$, while all remaining systems constitute the environment.
		All environmental states are assumed to be initially populated with equal weights, so that Eq.~\eqref{rho_time_1C} can be used}  
	\label{Fig:sample}
\end{figure}

\subsection{A numerical example of the exact dynamics}
\label{sec:numerics}

An explicit numerical realization of the microscopic dynamics is shown in Fig.~\ref{Fig:sample}, where we show the time evolution of these quantities for a system in which the environment $E$ consists of $5\times10^2$ two-level subsystems.
In the numerical simulation, all two-level systems, namely, the subsystems $A$ and $B$ and the environmental degrees of freedom, are randomly distributed within a sphere of unit radius (in arbitrary units).
All pairwise interactions are assumed to be proportional to the inverse of the relative distance between the particles, and the closest pair is identified as the subsystem $AB$.

To perform numerical simulations with a large environment, we assume that all environmental states are populated with equal weights, i.e., we assume $|f_k|=1/\sqrt{2^N}$.
Under this assumption, the sums over the $2^N$ terms appearing in Eq.~\eqref{rho_time_1A} factorize into products of $N$ cosine functions, yielding
\begin{eqnarray}
\label{rho_time_1C}
\Gamma_\alpha &=&  \frac{1}{2^N} \sum_k e^{-2 \imath \chi_k^\alpha t}
               = \prod_{j=1}^N \cos(\blue{2}\omega_{j\alpha} t), \\
\Lambda_\pm &=& \frac{1}{2^N} \sum_k  e^{-2 \imath (\chi_k^a \pm \chi_k^b)t}
               = \prod_{j=1}^N \cos[\blue{2}(\omega_{jA}\pm\omega_{jB})t] \nonumber,
\end{eqnarray}
where the relation between $\chi_k^\alpha$ and $\omega_{j\alpha}$ is given below Eq.~\eqref{eq:Ham_int_3}. 
This factorization significantly reduces the computational complexity compared to the general case.
\blue{The factor $2$ in the cosine arguments follows from the eigenvalue differences of the Pauli-$z$ interaction Hamiltonians written above.}

From the behavior of the purity shown in Fig.~\ref{Fig:sample}, two distinct dynamical regimes can clearly be observed.
These regimes can be traced back to a fundamental difference between the function $\Lambda_-$ and the other three functions appearing in Eq.~\eqref{rho_time_1A}, valid independently of the equal-population assumption used in the numerical computation.

To explain such a difference, we note that, since the environmental elements are randomly distributed, subsystems $A$ and $B$ see nearly the same environment.
As a consequence, the difference between $\chi_k^A$ and $\chi_k^B$, while generally nonzero, is expected to be much smaller than each of them individually, and even more so than their sum.
This implies that the evolution of $\Lambda_-$ occurs on a much longer timescale than that of $\Gamma_\alpha$ and $\Lambda_+$.

Physically, the emergence of distinct dynamical timescales can be naturally interpreted in terms of \emph{collective} and \emph{relative} decoherence channels.
At short times, decoherence is dominated by environmental fluctuations that act almost identically on subsystems $A$ and $B$.
This form of collective noise efficiently suppresses coherences associated with operators acting nontrivially on a single subsystem, while leaving essentially unaffected those coherences that depend on relative degrees of freedom between $A$ and $B$.

By contrast, the decay of $\Lambda_-$ is controlled by \emph{relative decoherence}, which originates from the small differences between the effective environments experienced by the two subsystems.
When $A$ and $B$ are spatially close and embedded in the same disordered many-body environment, these differences are weak, resulting in a strong suppression of the corresponding decoherence channel~\cite{Zanardi1997, Lidar1998}.
As a consequence, coherences associated with relative operators persist over much longer times.

This separation of timescales gives rise to an intermediate dynamical regime in which the functions $\Gamma_\alpha$ and $\Lambda_+$ can be considered as vanishing. 
In this regime, the reduced density matrix $\rho_{AB}(t)$ can be approximated by
\begin{equation}
\label{rho_time_lim}
\rho_{AB}(t)\simeq
\frac{1}{4}
\begin{pmatrix}
1 & 0 & 0 & 0 \\
0 & 1 & \Lambda_- & 0 \\
0 & \Lambda_-^* & 1 & 0 \\
0 & 0 & 0 & 1
\end{pmatrix}.
\end{equation}
At the onset of this regime, $\Lambda_- \simeq 1$ corresponding to a state with purity equal to $3/8$ and vanishing entanglement.
At longer times, however, the small but finite differences between $\chi_k^A$ and $\chi_k^B$ become dynamically relevant.
Although this process occurs on a much slower timescale, $\Lambda_-$ eventually decays as well, leading to the suppression of all remaining off-diagonal elements of the reduced density matrix.
\blue{In this late-time window, $\rho_{AB}(t)$ is approximately diagonal; this does not exclude finite-bath recurrences at later times.}

Before attempting to reproduce these behaviors through an effective Lindblad description, let us analyze their general features in detail.

\subsection{The short-time regime}

We begin with the short-time regime, where the exact dynamics can be treated analytically by expanding the functions entering the reduced density matrix in Eq.~\eqref{rho_time_1} in powers of time.
Specifically, we expand the quantities defined in Eq.~\eqref{rho_time_1A} up to second order in $t$, obtaining
\begin{eqnarray}
	\label{rho_time_1B1}
	\Gamma_\alpha &\simeq& \sum_k |f_k|^2 \left( 1 - 2\imath \chi_{k}^\alpha t - 2(\chi_k^\alpha)^2 t^2 \right), \nonumber \\
	\Lambda_\pm &\simeq& \sum_k |f_k|^2
	\left( 1 - 2\imath \chi_{k}^A t - 2(\chi_{k}^A)^2 t^2 \right) \times \\
	& & \qquad \times \left( 1 \mp 2\imath \chi_{k}^B t - 2(\chi_{k}^B)^2 t^2 \right), \nonumber
\end{eqnarray}
where terms of order $t^3$ and higher have been neglected.

This expansion makes explicit the role played by the statistical properties of the interaction strengths $\chi_k^\alpha$ in shaping the early-time dynamics.
To write these results in a compact form, it is convenient to introduce the weighted averages
\begin{equation}
	\mu_\alpha = \sum_k |f_k|^2 \chi_k^\alpha, \qquad
	\sigma_\alpha^2 = \sum_k |f_k|^2 (\chi_k^\alpha-\mu_\alpha)^2 ,
\end{equation}
which represent the mean values and variances of the distributions $\chi_k^\alpha$, with weights determined by the populations of the environmental states.
In addition, we define the covariance
\begin{equation}
	\sigma_c^2 = \sum_k |f_k|^2 (\chi_{A k}-\mu_A)(\chi_{B k}-\mu_B),
\end{equation}
which quantifies the correlations between the interaction strengths experienced by subsystems $A$ and $B$ weighted by the population of the environment's states.

For simplicity, and in line with the numerical setting discussed in the previous section, we assume that the two distributions are statistically equivalent, so that $\mu_A\simeq\mu_B=\mu$, $\sigma_A\simeq\sigma_B=\sigma$, and $\sigma_c^2\simeq\sigma^2$.
\blue{The relevant condition is $\operatorname{Var}(\chi^A-\chi^B)=\sigma_A^2+\sigma_B^2 -2\sigma_c^2 \ll\sigma^2$, as expected for a sufficiently close pair with similar couplings to the bath. 
A large environment alone does not guarantee this condition for arbitrary couplings or populations.}

The contribution associated with the mean value $\mu_\alpha$ can be interpreted as arising from a coherent unitary evolution generated by local effective fields induced by the environment and acting independently on subsystems $A$ and $B$.
In the present model, decoherence originates from the fact that the environment induces a superposition of distinct periodic evolutions of the subsystem $AB$ that are not perfectly phase aligned.
The resulting loss of coherence is therefore due to the progressive dephasing among these dynamical contributions.
By contrast, the terms proportional to $\mu_\alpha$ correspond precisely to the phase-coherent component of the environmental action, namely to the part of the evolution that remains fully in phase across all environmental realizations.
As such, they generate a deterministic unitary dynamics that can equivalently be described as the action of local effective fields proportional to $\sigma_\alpha^z$.
Since local unitary evolutions do not affect either the purity of the reduced state $\rho_{AB}(t)$ or the entanglement between $A$ and $B$, these contributions may be neglected for the purposes of the present analysis.

Retaining only the leading terms that give rise to nontrivial decoherence effects, one finds
\begin{equation}
	\label{rho_time_1B}
	\Gamma_\alpha \simeq e^{-2\sigma^2 t^2}, \qquad
	\Lambda_+ \simeq e^{-8\sigma^2 t^2}, \qquad
	\Lambda_- \simeq 1.
\end{equation}
Substituting these expressions into the density matrix in Eq.~\eqref{rho_time_1}, one finds that the purity of the reduced state decreases quadratically at short times, \blue{$P(t)\simeq 1-4\sigma^2 t^2$}, consistently with a Gaussian decay.
\blue{The coefficient follows directly from
\begin{equation}
P(t)=\frac14+\frac{|\Gamma_A|^2+|\Gamma_B|^2}{4} +\frac{|\Lambda_+|^2+|\Lambda_-|^2}{8}.
\label{eq:purity_factors}
\end{equation}
The quadratic onset is consistent with the qualitative behavior in the upper panel of Fig.~\ref{Fig:sample}.}
The deviations observed at longer times can be attributed to the increasing relevance of higher-order contributions beyond the second order in the time expansion.
As a result, the analytical approximation progressively departs from the exact dynamics, while preserving the overall qualitative structure and hierarchy of the dynamical regimes.

The same mechanism governs the behavior of the concurrence: while the oscillation frequency remains unaffected, its amplitude is progressively suppressed, reflecting the environmentally induced loss of coherence.

\subsection{The intermediate-time regime}

We now analyze the second dynamical regime, which emerges after the rapid suppression of the coherences associated with the fastest-decaying terms discussed in the previous subsection.
This regime is characterized by a clear separation of timescales.
\blue{The plateau at the onset of this regime lies in the temporal window} defined by
\begin{equation}
	\sigma^{-1} \ll t \ll \sigma_{\Lambda_-}^{-1},
\end{equation}

\blue{It is important to distinguish this regime from the initial short-time transient of the complete microscopic evolution. 
The lower bound $t\gg\sigma^{-1}$ implies that, upon entering this window, the fast collective channels governed by $\Gamma_\alpha$ and $\Lambda_+$ have already decayed and the reduced state has already departed substantially from its initial condition, reaching the intermediate plateau shown in Fig.~\ref{Fig:sample}. 
The condition $t\ll\sigma_{\Lambda_-}^{-1}$ therefore expresses short-time behavior only with respect to the second, parametrically slower relative-dephasing channel. 
The expansion below describes the onset of this distinct dynamical process, rather than the initial
short-time expansion of the full evolution.}

\blue{Here,} $\sigma$ controls the fast collective decoherence processes analyzed in the short-time regime, while $\sigma_{\Lambda_-}$ sets the much longer timescale associated with relative decoherence.
Within this time window, the matrix elements of the reduced density matrix that decay on the shortest timescale, namely $\Gamma_\alpha$ and $\Lambda_+$, can be approximated as vanishing. Hence, the reduced state of the subsystem $AB$ reduces to that of Eq.~\eqref{rho_time_lim},
where $\Lambda_-(t)$ is the only remaining time-dependent coefficient.
At the onset of this regime, the slow character of the relative decoherence channel implies $\Lambda_- \simeq 1$, while its subsequent evolution governs the long-time decay of the residual coherences.

To characterize the dynamics in this regime, we proceed as in the short-time analysis and perform a systematic expansion of $\Lambda_-$ in powers of time.
Recalling its definition in Eq.~\eqref{rho_time_1A}, we expand $\Lambda_-$ up to second order in $t$, obtaining
\begin{equation}
	\Lambda_- \simeq \sum_k |f_k|^2
	\left[
	1 - 2\imath (\chi_k^A-\chi_k^B)t
	- 2(\chi_k^A-\chi_k^B)^2 t^2
	\right],
\end{equation}
where higher-order terms have been neglected.

As in the previous regime, it is convenient to introduce the statistical moments of the distribution of frequency differences $\chi_k^A-\chi_k^B$:
\begin{eqnarray}
	\mu_{\Lambda_-} &=& \sum_k |f_k|^2 (\chi_k^A - \chi_k^B), \\
	\sigma_{\Lambda_-}^2 &=& \sum_k |f_k|^2 \big[(\chi_k^A-\chi_k^B)-\mu_{\Lambda_-}\big]^2. \nonumber
\end{eqnarray}
The contribution proportional to the mean value $\mu_{\Lambda_-}$ corresponds to a purely unitary evolution generated by an effective Hamiltonian acting locally on subsystems $A$ and $B$.
As discussed in the previous section, such local unitary dynamics leaves both the purity of the reduced state and the entanglement between the subsystems invariant and can therefore be neglected for the present analysis.
From a physical perspective, this term represents the coherent component of the relative decoherence channel, associated with phase-aligned contributions that do not induce dephasing.

\blue{Moreover, for uniformly populated environmental states at any finite $N$,} $|f_k|^2 = 2^{-N}$ for all $k$, the mean value $\mu_{\Lambda_-}$ vanishes identically.
\blue{This cancellation follows from the exact pairing of opposite environmental spin configurations in the uniformly weighted sum.}
As a result, only fluctuations around the mean survive, and the dynamics in this regime is entirely governed by the variance $\sigma_{\Lambda_-}^2$.

Retaining only the leading nontrivial contribution, one obtains
\begin{equation}
	\Lambda_-(t) \simeq \exp\!\left(-2\sigma_{\Lambda_-}^2\, t^2\right),
\end{equation}
\blue{which is the second-cumulant Gaussian approximation for the relative channel.
The Taylor expansion alone establishes only its quadratic onset.
For uniform environmental populations, its range of validity can be made explicit by setting $d_j=\omega_{jA}-\omega_{jB}$:
\begin{equation}
\begin{split}
\sigma_{\Lambda_-}^2&=\sum_j d_j^2,\\
\log\Lambda_-(t)&=-2t^2\sum_jd_j^2
+O\!\left(t^4\sum_jd_j^4\right).
\end{split}
\label{eq:slow_gaussian_control}
\end{equation}
This expansion follows from the exact cosine product and holds when $2t\max_j|d_j|\ll1$. 
If $\epsilon_-=\sum_jd_j^4/(\sum_jd_j^2)^2\ll1$, the Gaussian approximation can remain accurate for $t\sim\sigma_{\Lambda_-}^{-1}$, when the residual coherence changes appreciably, provided no individual $d_j$ dominates the variance. 
Combined with $\sigma_{\Lambda_-}\ll\sigma$, this gives a slow dephasing window after the fast channels have decayed.
The same argument applies to the fast channels with their respective coupling combinations. These conditions, rather than the Taylor expansion alone, justify using Gaussian envelopes beyond their onset.}

The evolution described by Eq.~\eqref{rho_time_lim}, together with the Gaussian decay of $\Lambda_-(t)$, leads to a slow and monotonic reduction of the purity from its plateau value $P=3/8$ toward \blue{the approximately decohered value} $P=1/4$.
Despite the presence of residual off-diagonal elements for finite times, the state remains separable throughout this regime, consistently with the vanishing concurrence observed in the numerical simulations.

\blue{Under the conditions stated above, the microscopic model exhibits two approximately Gaussian dephasing envelopes: a fast collective channel and a slower relative channel that remains active after the other coherences have been suppressed.} In the following sections, we introduce the effective non-unitary description of our system and assess whether the behaviors just discussed for two dynamical regimes can be faithfully reproduced within a Lindblad-type approach.

\section{The effective model}

Before addressing whether the Lindblad approach can reproduce the dynamics of the subsystem $AB$ within each of the regimes identified above, we briefly recall the standard framework of Markovian open quantum system dynamics used as an effective description of environmental effects.
Within this setting, the time evolution of the density matrix is described by the Gorini--Kossakowski--Sudarshan--Lindblad (GKSL) master equation~\cite{Gorini1976,Lindblad1976}.
For a bipartite system composed of subsystems $A$ and $B$, the evolution reads
\blue{\begin{equation}
\label{eq:GKSL}
 \frac{d}{dt}\rho_{AB}(t)
=
-i[H_{AB},\rho_{AB}(t)]
+
\mathcal{D}_{AB}[\rho_{AB}(t)]\blue{.}
\end{equation}}
where $\mathcal{D}_{AB}$ denotes the GKSL dissipator accounting for
the non-unitary contribution induced by the coupling to the
environment.

\blue{The purpose of the following construction is not to establish the general fact that constant-coefficient linear master equations lead to exponential relaxation. 
Instead, we use the operator structure inherited from the microscopic model to determine which sectors of the two-qubit density matrix must remain invariant and which sectors must decay in each dynamical regime. 
We then ask whether a time-homogeneous GKSL generator respecting this structure can reproduce both the pattern and the functional form of the exact decoherence.
A crucial feature of the time-homogeneous GKSL semigroup considered here is that its generator is time independent ~\cite{Lindblad1976, Gorini1976, Breuer2007}. 
The resulting dynamics forms a completely positive and trace-preserving quantum dynamical semigroup. In the pure-dephasing sectors relevant to the present model, constant dissipative rates therefore lead to
exponential coherence factors with fixed characteristic timescales~\cite{Breuer2007}.}

Of course, a crucial step is the choice of the appropriate Lindblad superoperator to be used in Eq.~\eqref{eq:GKSL} to describe the evolution of our system. For this, we shall consider the symmetries of the Hamiltonian and make use of the results of the previous section.

Instead of working directly with the reduced density matrix elements $\rho_{AB}^{j,l}$, $j,l=1 \ldots 4$, we find it convenient to express the GKSL master equation in the operator basis generated by tensor products of Pauli matrices acting on the two local two-level subsystems $A$ and $B$~\cite{Osborne2002}. We will denote their expectation values as
\begin{equation}
 \langle \sigma_A^{\mu}\otimes\sigma_B^{\nu} \rangle \equiv G^{\mu,\nu} (t) , \quad
 \mu,\nu=0,x,y,z \: .
\end{equation}

Since the Hilbert space of $AB$ has dimension $4$, the corresponding operator space is $16$-dimensional, and the Lindblad equation gives rise to a closed system of $16$ coupled first-order differential equations.
These equations can naturally be organized into four groups, each corresponding to a distinct sector of the reduced density matrix and characterized by a specific transformation of the local degrees of freedom.

The first group corresponds to the matrix elements lying on the main diagonal of the reduced density matrix $\rho_{AB}^{j,j}$ and is spanned by the operators associated to the correlation functions $G^{0,0}, G^{z,0}, G^{0,z}, G^{z,z}$.
All operators in this set commute with the full Hamiltonian in Eq.~\eqref{FullHamiltonian}, including both the internal Hamiltonian of the subsystem $AB$ and its interaction with the environment.
Their expectation values are therefore exact constants of motion under the microscopic unitary evolution.
Any effective Lindblad description aiming to reproduce the exact reduced dynamics must preserve these conservation laws.
\blue{The effective dissipator must annihilate these diagonal observables in the Heisenberg picture. This preserves populations without requiring the dephasing coefficients acting on off-diagonal sectors to vanish.} 

We next consider the matrix elements lying outside the two principal diagonals, which are characterized by a change in the degrees of freedom of one subsystem while those of the other remain unchanged, namely, the elements $\rho_{AB}^{1,2}, \rho_{AB}^{2,1}$ and $\rho_{AB}^{1,3}, \rho_{AB}^{3,1}$.
In the former set, the state of subsystem $B$ is modified while that of subsystem $A$ remains unchanged, whereas in the second set the role of $A$ and $B$ is interchanged. 
\blue{We discuss the sector that changes $A$, remembering that analogous considerations apply to the sector that changes $B$. In this sector,} the 4 relevant operators can be divided into two pairs, which are dynamically coupled. Namely, the operators in $G^{x,0}$ produce those in $G^{y,z}$ upon commutation with $H_{AB}$ from Eq.~\eqref{HAB_2} and vice versa. Similarly for $G^{y,0}$ and $G^{x,z}$. Moreover, since at $t=0$ only $G^{x,0}$ is nonvanishing, while all other correlation functions in this sector vanish identically, the dynamics associated with the second coupled pair remains trivial and can be consistently neglected.
The analysis can therefore be restricted to the first coupled pair.

Finally, we consider the matrix elements associated with the anti-diagonal of the reduced density matrix $\rho_{AB}^{j,5-j}$, which involve operators that simultaneously modify the degrees of freedom of both subsystems.
The corresponding set is given by $G^{x,x}, G^{x,y}, G^{y,x}, G^{y,y}$.
Although all the operators connected with these expectation values commute with the Hamiltonian $H_{AB}$, their expectation values are not conserved under the full microscopic dynamics, since they do not commute with the interaction Hamiltonians $H_{AE}$ and $H_{BE}$.
As a consequence, this sector exhibits a genuinely nontrivial evolution.

Having established the assumptions and structure underlying the Lindblad framework, we now analyze the dynamics in the two different temporal regimes, with the explicit aim of assessing whether and to what extent the exact reduced dynamics, specifically the evolution of entanglement between $A$ and $B$ and the purity of $\rho_{AB}(t)$, can be faithfully reproduced within a time-homogeneous Lindblad description.

\subsection{The short-time regime}

\blue{The corresponding effective generator can be derived explicitly from a two-qubit pure-dephasing GKSL equation with a positive Kossakowski matrix. 
The complete derivation is given in Appendix~\ref{SEC_APP_C}. Here we only report the resulting equations for the
relevant operator sectors. 
In the short-time regime, the Kossakowski coefficients can be chosen so as to represent fully correlated dephasing of the two subsystems. 
For the sector spanned by $G^{x,0}$ and $G^{y,z}$, this gives}

\blue{\begin{align}
\label{eq:GKSL_off}
\frac{d}{dt}G^{x,0} &=2\omega G^{y,z}-\frac{\lambda}{2}G^{x,0},
\\
\frac{d}{dt}G^{y,z} &=-2\omega G^{x,0}-\frac{\lambda}{2}G^{y,z}.
\end{align}}
The corresponding solutions read
\blue{\begin{align}
\label{eq:GKSL_4}
G^{x,0}(t) &=e^{-\lambda t/2}\cos(2\omega t),
\\
G^{y,z}(t) &=-e^{-\lambda t/2}\sin(2\omega t).
\end{align}}

At this stage, an important difference between the exact dynamics and its Lindblad reconstruction already emerges.
While the microscopic model predicts a Gaussian suppression of coherences at short times, with leading terms proportional to $t^2$, the Lindblad evolution necessarily produces an exponential decay linear in time.
This mismatch is not a quantitative detail but a structural consequence of the time-homogeneous, semigroup nature of the GKSL generator.
As we show in the following, this discrepancy becomes even more pronounced when analyzing the remaining sectors of the reduced density matrix and ultimately prevents a faithful Lindblad description of the short-time dynamics.

Turning now to the anti-diagonal components, in agreement with the numerical results and the analytical arguments presented in the previous sections, in the short-time regime the two coefficients $\Lambda_\pm$ entering the reduced density matrix display markedly different behaviors.
While $\Lambda_-\simeq 1$ and therefore remains effectively constant, $\Lambda_+$ retains a nontrivial time dependence.
Taking the structure of the initial state in Eq.~\eqref{initialstate} into account, which implies that at $t=0$ only the expectation value $G^{x,x}$ is nonvanishing, the effective Lindblad dynamics in this sector reduces to a pair of coupled differential equations, which we can write as:
\blue{\begin{align}
\label{eq:GKSL_sec} \frac{d}{dt}G^{x,x} &=-\phi G^{x,x}+\delta G^{y,y},
\\
\frac{d}{dt}G^{y,y} &=-\phi G^{y,y}+\delta G^{x,x}.
\end{align}}

For the initial conditions imposed by the chosen state, this system has the solution
\blue{\begin{align}
\label{eq:GKSL_4_1}
G^{x,x}(t)
&=e^{-\phi t}\cosh(\delta t),
\\
G^{y,y}(t)
&=e^{-\phi t}\sinh(\delta t).
\end{align}}

By imposing $\phi=\delta=\lambda$, one obtains the Lindblad functional form
\begin{equation}
\label{rho_time_1c}
\Gamma_\alpha \simeq e^{-\lambda t/2}, \qquad
\Lambda_+ \simeq e^{-2\lambda t}, \qquad
\Lambda_- \simeq 1.
\end{equation}
At first sight, these expressions appear very similar to those obtained within the microscopic model.
Indeed, all matrix elements that are time independent in the short-time regime of the microscopic dynamics remain time independent within the Lindblad description, and the same correspondence holds for the time-dependent ones.
This apparent agreement, however, is only superficial.

Indeed, a crucial difference emerges when comparing the short-time expansions of the two descriptions.
\blue{After removing the coherent local phases, the microscopic coherence magnitudes have a leading decay proportional to $t^2$. The Gaussian approximations used above are even functions of time.}
By contrast, the corresponding functions generated by the Lindblad dynamics are not even functions of time, as their expansion around the initial time contains terms linear in $t$.
This mismatch is not a quantitative detail, but a direct consequence of the above-mentioned time-homogeneous, semigroup structure underlying the GKSL generator.

Importantly, this discrepancy cannot be removed by tuning the Lindblad parameters.
\blue{For the pure-dephasing sectors considered here, any nonzero time-independent dephasing rate in the GKSL generator produces exponential coherence factors and therefore a nonvanishing term linear in $t$ in their short-time expansion. 
This differs from the quadratic behavior of the exact microscopic coherence factors. Importantly, this statement concerns a time-homogeneous GKSL semigroup with constant coefficients. A time-local master equation with explicitly time-dependent rates can reproduce Gaussian dephasing and is not excluded by the present analysis. 
For example, $\lambda(t)=8\sigma^2t\geq0$ reproduces the short-regime Gaussian factors in Eq.~\eqref{rho_time_1B} with a positive instantaneous Kossakowski matrix. 
Thus even CP-divisible time-dependent dynamics is not excluded by these Gaussian factors.}

\subsection{The intermediate-time regime}

We now attempt to reproduce the dynamics of the subsystem $AB$ in this second temporal regime within a Lindblad framework.
Since the reduced density matrix is now effectively described by Eq.~\eqref{rho_time_lim}, only the diagonal elements and the off-diagonal matrix elements $\rho_{AB}^{2,3}, \rho_{AB}^{3,2}$ on the anti-diagonal remain dynamically relevant.
As in the short-time regime, the diagonal elements correspond to operators that commute with the full microscopic Hamiltonian and must therefore remain constants of motion.
\blue{Consistency with the exact dynamics again requires an invariant population sector, while off-diagonal dephasing rates may remain nonzero.}

We thus focus on the nontrivial sector associated with the coherences $\rho_{AB}^{2,3}, \rho_{AB}^{3,2}$, which are governed by the coefficient $\Lambda_-(t)$.
Formally, this situation mirrors the role played by $\Lambda_+(t)$ in the short-time regime, with the crucial difference that decoherence now acts on elements closer to the diagonal rather than on $\rho_{AB}^{1,4}, \rho_{AB}^{4,1}$ sector.

\blue{The same Kossakowski construction can be used in this regime, but with anticorrelated rather than correlated dephasing channels, as shown explicitly in Appendix~\ref{SEC_APP_C}. 
Choosing the correlation parameter with the opposite sign interchanges the decay rates of the two anti-diagonal coherences. 
In particular, one can choose the constant coefficients such that the already suppressed  $\Lambda_+$ coherence remains zero while $\Lambda_-(t)=\Lambda_-(t_0)e^{-\tilde{\lambda}(t-t_0)}$, where $t_0$ is a time at which the intermediate-regime description is initialized. 
Thus, the time-homogeneous GKSL model can reproduce which residual coherence decays in the intermediate regime, but it still predicts an exponential rather than the Gaussian dependence obtained from
the microscopic dynamics.
}

Within this construction, the Lindblad dynamics reproduces the qualitative structure of the exact evolution in the intermediate-time regime, namely a slow and monotonic decay of the surviving off-diagonal elements together with constant populations.
However, as in the short-time case, this agreement remains only qualitative.
While the Lindblad equation can be tuned to reproduce which matrix elements decay and which remain constant, it fails to capture the correct functional form of the decay.
\blue{Within the Gaussian approximation established above, a comparison that uses the same initial coherence at $t_0$ gives
\begin{equation}
\frac{\Lambda_-(t)}{\Lambda_-(t_0)} \simeq e^{-2\sigma_{\Lambda_-}^2(t^2-t_0^2)}
\end{equation}
for the microscopic model, whereas the effective generator yields $e^{-\tilde{\lambda}(t-t_0)}$. 
The corresponding logarithmic decay rates are $4\sigma_{\Lambda_-}^2t$ and $\tilde{\lambda}$, respectively. 
A constant rate cannot reproduce the Gaussian envelope throughout this window, although an exponential fit over a sufficiently narrow interval is not excluded.}

\blue{The mismatch therefore persists beyond the initial short-time transient of the full microscopic evolution and reappears at the onset of the parametrically slower relative-dephasing channel.
This shows that the discrepancy is not confined to the initial transient associated with the fast collective channel.}

\section{Conclusions}

In this work, we analyzed, in a fully controlled setting, the extent to which a Lindblad master equation can faithfully reproduce the reduced dynamics emerging from a microscopic many-body quantum system.
Focusing on two interacting two-level subsystems embedded in a larger environment and coupled exclusively through unitary, pairwise interactions, we compared a first-principles microscopic treatment with an effective description based on a time-homogeneous Gorini--Kossakowski--Sudarshan--Lindblad (GKSL) generator.

The exact microscopic dynamics exhibits a clear separation between dynamical regimes.
At short times, decoherence arises from the dephasing among distinct unitary evolutions induced by the environment, leading to \blue{an approximately Gaussian} suppression of coherences and a quadratic decay of the purity.
At intermediate times, a second regime emerges in which collective decoherence channels are effectively saturated, while a much slower decay driven by relative environmental fluctuations governs the remaining off-diagonal elements of the reduced density matrix. \blue{Toward the end of this second regime, the remaining coherences are strongly suppressed in the observed time window; finite-bath recurrences are not excluded.}
\begin{table}[t!]
	\begin{center}
			\begin{tabular}{| c |c | c |} 
				\hline
				Regime & Unitary & Effective \\ [0.5ex] 
				\hline
				\hline
				Short & & \\
				$\Gamma_\alpha \simeq$  & $e^{-2\sigma^2 t^2}$ & $e^{-\lambda t/2}$\\ 
				$\Lambda_+ \simeq $ & $e^{-8\sigma^2 t^2}$& $e^{-2\lambda t}$  \\
				$\Lambda_- \simeq$ & $1$ &  $1$ \\
				\hline
				\hline
				Intermediate & & \\
				$\Gamma_\alpha \simeq$  & $0$ &  $0$ \\ 
				$\Lambda_+ \simeq $ & $0$&  $0$  \\
				$\Lambda_- \simeq$ & $e^{-2\sigma_{\Lambda_-}^2 t^2}$ &  $\Lambda_-(t_0) e^{-\tilde{\lambda} (t-t_0)}$  \\
				\hline
				\hline
				Late & & \\
				$\Gamma_\alpha \simeq$  & $0$ & $0$  \\ 
				$\Lambda_+ \simeq $ & $0$& $0$  \\
				$\Lambda_- \simeq$ & $0$ & $0$  \\
				\hline
			\end{tabular}
			\caption{\label{table1} Recap of the behaviors of the reduced density matrix coefficients in Eq.~\eqref{rho_time_1}, for the short, intermediate, and Late regimes, according to the microscopic, unitary dynamics and the effective Lindblad description. \blue{The microscopic Gaussian envelopes are approximations under the conditions specified in Sec.~III, whereas the time-homogeneous pure-dephasing generators give exponential factors. The late window refers to times before recurrences.}}
	\end{center}
\end{table}

\blue{The microscopic dynamics admits Gaussian dephasing approximations in both regimes under the conditions discussed in Sec.~III.}
We have shown that, although a Lindblad master equation can be tuned phenomenologically to reproduce which matrix elements decay and which remain constant in each regime, it systematically fails to capture the correct functional form of the decay.
\blue{In particular, for the pure-dephasing semigroup compatible with the coherence sectors of the present microscopic model, any nonzero constant dephasing rate produces a linear-in-time contribution in the short-time expansion of the corresponding coherence factor, in contrast with the quadratic behavior of the exact microscopic dynamics. 
This mismatch cannot be removed by retuning the constant dephasing rates within this time-homogeneous construction.}
Our analysis further demonstrates that this limitation persists even when the Lindblad description is adapted separately to different dynamical regimes.
\blue{A piecewise construction with different generators in the two regimes does not define a single time-homogeneous semigroup for the full evolution.
Thus, in the present model the mismatch is not removed by adapting the constant dephasing rates separately to the two dynamical regimes. 
Rather, it reflects the incompatibility of the two-scale microscopic coherence factors with a single time-homogeneous pure-dephasing semigroup.}

\blue{The Gaussian coherence factors obtained from the microscopic evolution should not, by themselves, be interpreted as a criterion for non-Markovianity. 
What they establish here is the failure of a single time-homogeneous semigroup with constant dephasing rates to reproduce the exact reduced dynamics. This distinction is important: time-dependent time-local generators may reproduce Gaussian dephasing, and the absence of a time-homogeneous semigroup does not
automatically imply information backflow. 
As discussed in Appendix~\ref{SEC_APP_B}, the trace-distance analysis performed for the realizations considered here does not reveal such a backflow for the sampled initial-state pairs and time interval; this finite numerical test does not prove its absence for all states or later times.}

\blue{The solvable model provides a controlled benchmark for collective and relative pure dephasing in a bipartite subsystem. Its non-decaying bath correlations explain why the usual Born--Markov construction is not applicable. The comparison does not establish a general limitation of GKSL equations, nor does it rule out a coarse-grained exponential approximation over a restricted interval.}

The present work naturally suggests several directions for future investigation.
First, it would be interesting to extend the analysis to microscopic models involving non-commuting interaction Hamiltonians and genuine energy-exchange processes, in order to assess to what extent the structural limitations identified here persist beyond pure-dephasing scenarios.
Second, a systematic comparison with time-dependent or explicitly non-Markovian master equations could clarify under which conditions such generalized approaches provide a faithful and physically consistent description of the reduced dynamics.
Finally, exploring experimentally accessible signatures that distinguish Gaussian from exponential decoherence, particularly in platforms involving many-body environments or gravitationally induced entanglement, would represent a natural step toward connecting the present theoretical results with ongoing experimental efforts, and toward assessing the reliability of effective open-system descriptions in complex quantum settings.

\acknowledgements{FF and SMG acknowledge support from the project "Implementation of cutting-edge research and its application as part of the Scientific Center of Excellence for Quantum and Complex Systems, and Representations of Lie Algebras", Grant No. PK.1.1.10.0004, co-financed by the European Union through the European Regional Development Fund-Competitiveness and Cohesion Programme 2021-2027. Moreover, FF acknowledges the support of the Croatian Science Foundation (HrZZ) through the project IP-2025-02-1667, Mining the Quantum: Frustration, Disorder, and Devices.}

\appendix

\section{Reduced density matrix}
\label{SEC_APP_A}

In this Appendix, we provide the explicit derivation of the analytical expression for the reduced density matrix given in Eq.~\eqref{rho_time_1}.  
We begin by recalling the main assumptions of our framework.  
At $t=0$, the global system is assumed to be in a fully factorized state,  
$\ket{\psi(0)} = \ket{\psi_A} \ket{\psi_B} \ket{\psi_e}$,
where the subsystem states $\ket{\psi_A}$ and $\ket{\psi_B}$ are defined in Eq.~\eqref{initialstate}, and the environmental state $\ket{\psi_e}$ is specified in Eq.~\eqref{psieexp}.  
Moreover, the system evolves under the Hamiltonian introduced in Eq.~\eqref{FullHamiltonian}.
Under these assumptions, the time-evolved state at any $t>0$ can be written as 
\begin{eqnarray} \label{tstate}
 \ket{\psi(t)} &=& e^{-iHt}\ket{\psi(0)} \\
 &=& \frac{1}{2}\sum_{ijk} f_k \, e^{-i(g_{ij}+h^a_{ik}+h^b_{jk}+\epsilon_k)t} \ket{a_i b_j e_k}, \nonumber
\end{eqnarray}
\blue{where $h_{ik}^{\alpha}=(-1)^{i+1}\chi_k^\alpha$, with $i=1,2$ and $\alpha=a,b$. 
From this expression we obtain the following density matrix:}
\begin{eqnarray}
    \rho(t) &=& \ket{\psi(t)}\!\!\bra{\psi(t)} \\
     &=&  \frac{1}{4}\sum_{\substack{ijk \\ i'j'k'}} f_k f^*_{k'} 
     \exp\left[-\imath \mu_{ijk}^{i',j',k'}t\right] 
     \ket{a_i b_j e_k}\bra{a_{i'} b_{j'} e_{k'}}. 
    \nonumber
\end{eqnarray}
Here, the frequency difference is defined as
\begin{equation}
\mu_{ijk}^{i',j',k'} = (g_{ij}-g_{i'j'}) + (h^a_{ik}-h^a_{i'k'}) + (h^b_{jk}-h^b_{j'k'}) + (\varepsilon_k-\varepsilon_{k'}), \nonumber
\end{equation}
where $\varepsilon_k$ denotes the eigenvalue of the environmental Hamiltonian $H_E$ associated with the eigenstate $\ket{e_k}$, $H_E\ket{e_k}=\epsilon_k\ket{e_k}$.

\begin{figure}[t!]
	\centering
	\includegraphics[width=\columnwidth]{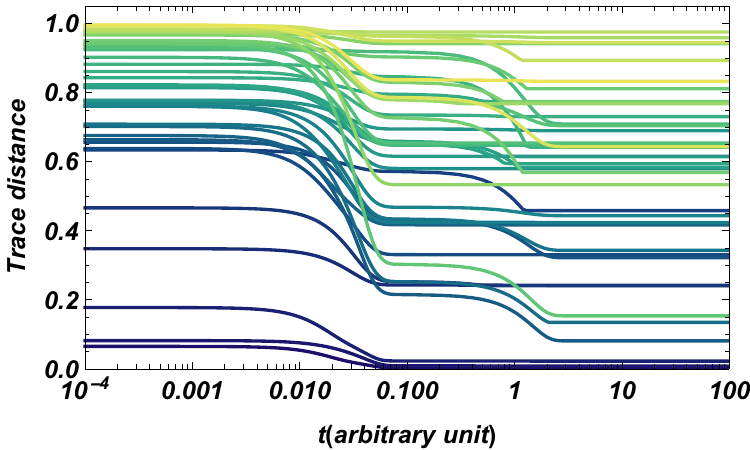}
	\caption{We consider the unitary evolution of two random initial states of subsystem $AB$ in a fixed environment, governed by the Hamiltonian in Eq.\eqref{FullHamiltonian}. We plot 40 time evolutions of the trace distance between the reduced density matrices of $AB$. \blue{No trace-distance increase is resolved for these sampled pairs in the plotted interval; this does not establish Markovianity for all initial states or times.}}
	\label{Fig:TraceDistance}
\end{figure}

To obtain the reduced density matrix for subsystems $A$ and $B$, we trace out the environmental degrees of freedom:
\begin{equation}\begin{split}\small
    &\rho_{AB}(t) = \sum_{k''}\bra{e_{k''}} \rho(t) \ket{e_{k''}}=\\
     &=  \frac{1}{4}\sum_{\substack{ijk  i'j'k'}} f_k f^*_{k} 
    \exp\left[-\imath \Delta_{ijk}^{i',j'}t\right] 
    \ket{a_i b_j}\bra{a_{i'} b_{j'}}\delta_{kk'}\delta_{kk''}=\\
    &= \frac{1}{4}\sum_{\substack{ijk \\ i'j'}} |f_k|^2  
    \exp\left[-\imath \Delta_{ijk}^{i',j'}t\right] 
    \ket{a_i b_j}\bra{a_{i'} b_{j'}}
    \end{split}\nonumber
\end{equation}
where
\begin{equation}
  \Delta_{ijk}^{i',j'} = (g_{ij}-g_{i'j'}) + (h^a_{ik}-h^a_{i'k}) + (h^b_{jk}-h^b_{j'k}).
\end{equation}
\blue{After removing the local fields and global phase as in Sec.~II, one has $g_{11}=g_{22}=-\omega$ and $g_{12}=g_{21}=\omega$. Together with $(h^\alpha_{1k}-h^\alpha_{2k})=2\chi_k^\alpha$, this gives the phases and coherence factors in Eq.~\eqref{rho_time_1}.}

\section{Tests of Markovianity}
\label{SEC_APP_B}

\blue{The unitary evolution of the total system does not by itself determine the Markovianity of the reduced subsystem. Here we test for information backflow in the reduced $AB$ dynamics.} We employ the results of 
\cite{Breuer2009,Breuer2012,Rivas2010} and consider the trace distance $D(\rho_1,\rho_2) = \frac{1}{2} {\rm tr} | \rho_1 - \rho_2 |$ between two reduced density matrices $\rho_{1}$ and $\rho_2$, where $|A| \equiv \sqrt{A^\dagger A}$. In \cite{Breuer2009} it was shown that in a Markovian system it is not possible to choose two initial states for which their trace distance increases in time. Thus, we follow the evolution under the unitary dynamics of pairs of reduced density matrices with random initial conditions for the $AB$ subsystem. The environment is instead kept fixed and chosen as in the numerical analysis discussed in Sec.\ref{sec:numerics}. \blue{Fig.~\ref{Fig:TraceDistance} shows no resolved increase of the trace distance for the sampled pairs in the plotted time interval. Since the test does not optimize over all initial states or cover arbitrarily long times, it cannot prove the absence of information backflow or CP-divisibility.}

\section{\blue{Explicit construction of the effective GKSL generator}}
\label{SEC_APP_C}

\blue{The effective equations used in Sec.~IV can be obtained explicitly
from a two-qubit pure-dephasing GKSL generator. 
This is a phenomenological construction, not a Born--Markov derivation from the microscopic bath with non-decaying correlations. It provides an admissible population-preserving generator, without asserting uniqueness.
We write the dissipative part of the master equation in the standard Kossakowski form
\begin{equation}
\mathcal{D}[\rho]  \sum_{\alpha,\beta=A,B} C_{\alpha\beta} \left[ \sigma_\alpha^z \rho \sigma_\beta^z
-\frac{1}{2} \left\{ \sigma_\beta^z\sigma_\alpha^z,\rho \right\} \right],
\end{equation}
where the Kossakowski matrix $C$ is positive semidefinite. 
Since we are interested in correlated pure dephasing, we take
\begin{equation}
C = c \begin{pmatrix}
1 & \eta\\
\eta & 1
\end{pmatrix},
\qquad
c\geq0,
\qquad
|\eta|\leq1.
\end{equation}
The positivity condition $|\eta|\leq1$ guarantees complete positivity of the corresponding dynamical semigroup.}

\blue{Let $|s\rangle$ denote one of the computational-basis states of the two-qubit subsystem and define
\begin{equation}
\sigma_\alpha^z|s\rangle = z_\alpha(s)|s\rangle,
\qquad
z_\alpha(s)=\pm1 .
\end{equation}
For an off-diagonal matrix element $\rho_{ss'}=\langle s|\rho|s'\rangle$, the dissipative contribution
is then
\begin{equation}
\left.
\frac{d}{dt}\rho_{ss'}
\right|_{\mathcal D}
=
-\frac{1}{2}
\sum_{\alpha,\beta=A,B}
C_{\alpha\beta}
\Delta z_\alpha
\Delta z_\beta
\,\rho_{ss'},
\end{equation}
where
\begin{equation}
\Delta z_\alpha
=
z_\alpha(s)-z_\alpha(s').
\end{equation}}

\blue{For coherences that differ only in one subsystem, such as $\rho_{12}$ or $\rho_{13}$, one has
\begin{equation}
\gamma_{\rm single}=2c.
\end{equation}
For the two anti-diagonal coherences one instead obtains
\begin{align}
\gamma_{14}
&=
4c(1+\eta),
\\
\gamma_{23}
&=
4c(1-\eta).
\end{align}
Since $\rho_{14}$ is proportional to $\Lambda_+$ and $\rho_{23}$ to $\Lambda_-$, the role of correlated and anticorrelated dephasing becomes transparent.}

\blue{In the first dynamical regime we choose fully correlated noise, $\eta=1$, and set
\begin{equation}
c=\frac{\lambda}{4}.
\end{equation}
It follows that
\begin{equation}
\gamma_{\rm single}
=
\frac{\lambda}{2},
\qquad
\gamma_{14}=2\lambda,
\qquad
\gamma_{23}=0,
\end{equation}
and therefore
\begin{equation}
\Gamma_\alpha(t)
=
e^{-\lambda t/2},
\qquad
\Lambda_+(t)
=
e^{-2\lambda t},
\qquad
\Lambda_-(t)
=
1.
\end{equation}
This reproduces Eq.~\eqref{rho_time_1c}.}

\blue{The Hamiltonian contribution generated by
$H_{AB}=-\omega\sigma_A^z\otimes\sigma_B^z$ couples the operators
$\sigma_A^x\otimes\sigma_B^0$ and
$\sigma_A^y\otimes\sigma_B^z$. Using
\begin{equation}
i[H_{AB},\sigma_A^x\otimes\sigma_B^0]
=
2\omega\,\sigma_A^y\otimes\sigma_B^z
\end{equation}
and
\begin{equation}
i[H_{AB},\sigma_A^y\otimes\sigma_B^z]
=
-2\omega\,\sigma_A^x\otimes\sigma_B^0,
\end{equation}
one obtains
\begin{align}
\dot G^{x,0}
&=
2\omega G^{y,z}
-\frac{\lambda}{2}G^{x,0},
\\
\dot G^{y,z}
&=
-2\omega G^{x,0}
-\frac{\lambda}{2}G^{y,z},
\end{align}
whose solutions are those reported in Sec.~IV A.}

\blue{For the anti-diagonal sector, the combinations
\begin{equation}
G^{x,x}-G^{y,y}
\end{equation}
and
\begin{equation}
G^{x,x}+G^{y,y}
\end{equation}
are proportional, respectively, to the real parts of $\rho_{14}$ and $\rho_{23}$. 
Their decay rates are therefore $\gamma_{14}$ and $\gamma_{23}$. 
Equivalently,
\begin{align}
\dot G^{x,x}
&=
-4c\,G^{x,x}
+
4c\eta\,G^{y,y},
\\
\dot G^{y,y}
&=
-4c\,G^{y,y}
+
4c\eta\,G^{x,x}.
\end{align}
For $\eta=1$ and $c=\lambda/4$, this gives $\phi=\delta=\lambda$, reproducing Eqs.~\eqref{eq:GKSL_sec} and \eqref{eq:GKSL_4_1} and hence the coherence factors in Eq.~\eqref{rho_time_1c}.}

\blue{In the intermediate regime, the opposite choice $\eta=-1$ interchanges the two anti-diagonal decay channels. 
Choosing
\begin{equation}
c=\frac{\tilde{\lambda}}{8}
\end{equation}
gives
\begin{equation}
\gamma_{14}=0,
\qquad
\gamma_{23}=\tilde{\lambda},
\end{equation}
so that
\begin{equation}
\blue{\Lambda_+(t)=\Lambda_+(t_0)\simeq0},
\qquad
\blue{\Lambda_-(t)=\Lambda_-(t_0)e^{-\tilde{\lambda}(t-t_0)}}.
\end{equation}
Here $t\geq t_0$, and $\Lambda_-(t_0)\simeq1$ near the plateau. 
In the intermediate regime the $\Lambda_+$ sector has already vanished as a consequence of the preceding fast dynamics; the effective generator therefore acts only on the surviving $\Lambda_-$ coherence. 
This explicitly shows that the time-homogeneous GKSL construction can reproduce the hierarchy of decaying sectors in the two regimes, while the exact microscopic dynamics differs in the functional form of the corresponding coherence factors.}

\bibliographystyle{apsrev4-2}
\bibliography{bibliografia}

\end{document}